\documentclass{article}

\usepackage{arxiv}

\usepackage[utf8]{inputenc} 
\usepackage[T1]{fontenc}    
\usepackage{hyperref}       
\usepackage{url}            
\usepackage{booktabs}       
\usepackage{amsfonts}       
\usepackage{nicefrac}       
\usepackage{microtype}      
\usepackage{lipsum}
\usepackage{graphicx}

\usepackage{amsmath}
\usepackage{orcidlink}
\usepackage{algorithm}        
\usepackage{algpseudocode}    
\usepackage{amssymb}          
 
\graphicspath{ {./images/} }

\title{Efficient DFT of Zadoff-Chu Sequences using lmFH Pattern}
\author{
  Fanping Du \orcidlink{0009-0002-9681-5521} \\
  CmosTek Microelectronics Co., Ltd.\\
  Shenzhen, China 518052 \\
  \texttt{1511445443@qq.com} \\
  }
\begin{document}
\maketitle
\begin{abstract}
Having established that Zadoff-Chu (ZC) sequences are inherently linear micro-frequency hopping (lmFH) symbols, this paper first presents an intuitive and visual exposition of the computation of the DFT and IDFT of ZC sequences using the lmFH pattern. This yields interesting results. Subsequently, an alternative form for computing the cumulative sum of ZC sequences using the Generalized Quadratic Gauss Sum is introduced. Furthermore, building on the micro-frequency hopping (mFH) concept, this paper shows that the DFT of ZC sequences can be transformed into an lmFH symbol with frequency shift and phase offset. Therefore, the DFT of ZC sequences can be computed via cumulative frequency points, similar to the computation of normal mFH symbols.
\end{abstract}
\section{Introduction}
A novel spread-spectrum modulation technique, termed micro-frequency hopping (mFH), has revealed that Zadoff-Chu (ZC) sequences can be interpreted as linear micro-frequency hopping (lmFH) symbols, and thus ZC sequences can also be intuitively and visually represented by the lmFH pattern, as illustrated in Fig. 4 of Ref. \cite{1}. As is well known, ZC sequences possess several distinctive properties, such as perfect autocorrelation and cross-correlation in both the time and frequency domains \cite{2}. Moreover, when ZC sequences have prime length, the Discrete Fourier Transform (DFT) of the ZC sequences is a conjugated, scaled, and time-scaled version of itself. This property has improved the computational efficiency of the DFT of ZC sequences greatly\cite{3}. However, by employing the mFH concept, the computational efficiency of the DFT of ZC sequences can be significantly further enhanced from a completely new perspective.
\section{Understanding ZC sequences using lmFH pattern}
\begin{figure}[H] 
	\centering
	\includegraphics[width=7cm]{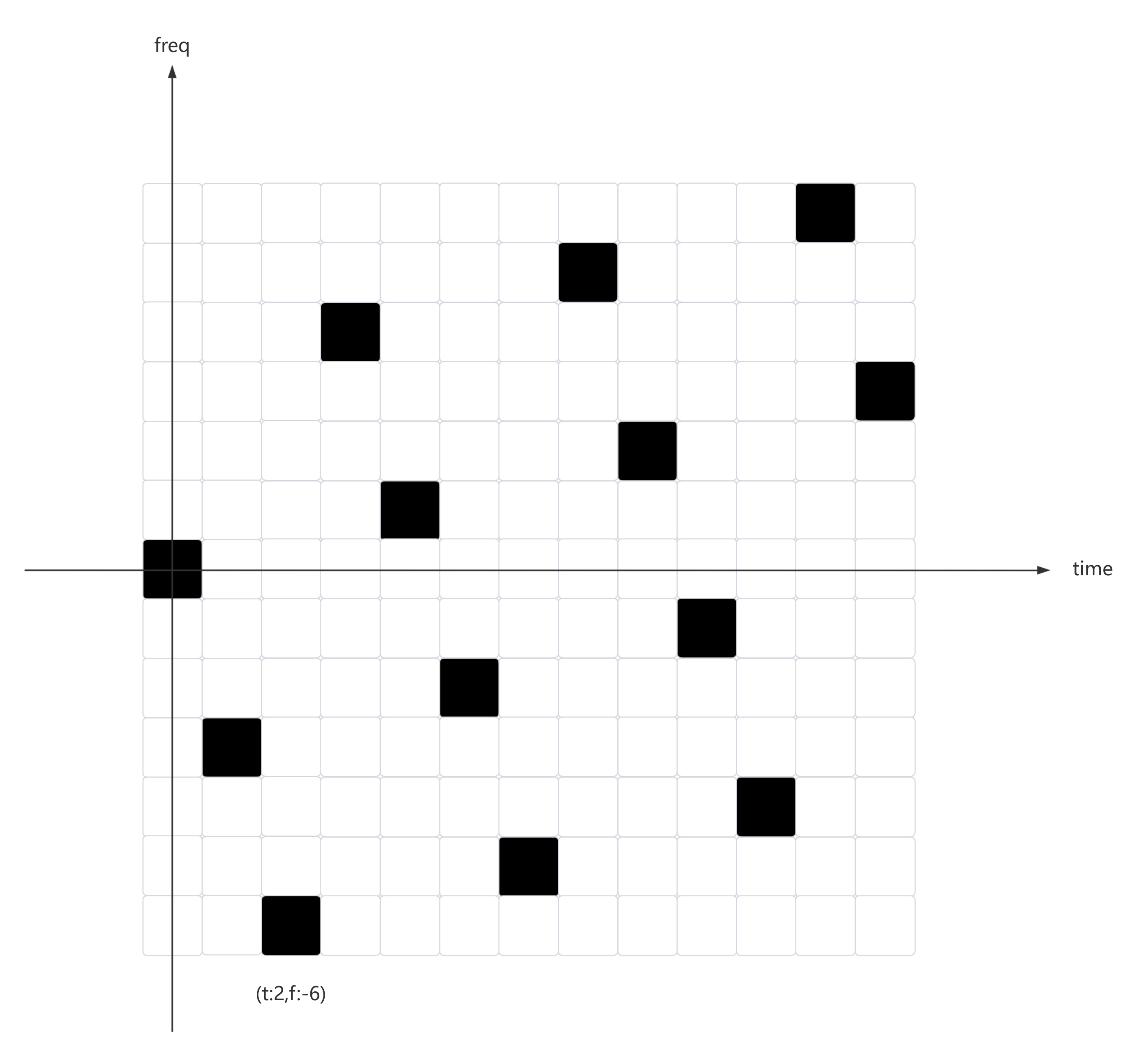}
	\caption{lmFH pattern with P13 \& u3}
	\label{fg:lmFH_P13_u3}
\end{figure}

ZC sequences are lmFH symbols in nature. According to the definition of the mFH symbol, where the phase is determined by the cumulative sum of frequency points in the mFH pattern, an lmFH symbol with a slope \(s\), \(s \not\equiv 0 \pmod P\), and a prime length \(P\) can be expressed as:
\begin{equation} \label{eq:lmFH}
L_{s}(k) = \exp(i 2\pi \frac{\sum_{t=0}^{k} s \cdot t}{P}) 
\end{equation}
where \(L_s\) is the lmFH symbol with slope \(s\), \(\exp(\cdot)\) is the exponential function, \(i=\sqrt{-1}\), \(t\) are time points in the lmFH pattern, the frequency points \(f\) in the lmFH pattern are \(f = s \cdot t \pmod P\), and \(k=0, 1, \dots, P-1\). \\ 
It is evident that the lmFH symbol is the conjugate of the ZC sequence, i.e., \(L_s = Z_u^*\), or \(L_{-s} = Z_u\), meaning that the ZC sequences with root \(u\) corresponds to an lmFH pattern with slope of \(-s\). Therefore, we use \(-u\) to denote the slope of the lmFH pattern. Fig. \ref{fg:lmFH_P13_u3} shows an example of an lmFH pattern with length \(P=13\) and slope \(s=-u=-3\). \\
Furthermore, according to the definition of micro-frequency hopping cyclic frequency shift (mFHCFS) modulation \cite{1}, an lmFH symbol with a frequency shift \(F_s\) can be expressed as:
\begin{equation} \label{eq:lmFH_Fs}
L_{-u}(k) = \exp(i 2\pi \frac{\sum_{t=0}^{k} (-u \cdot t + F_s')}{P}) 
\end{equation}
Note: Here, \(F_s' = F_s\) when \(t \neq 0\); otherwise, \(F_s' = 0\) to avoid introducing an extra initial phase. Unless otherwise specified, in this paper, when expressing ZC sequences in the form of lmFH symbols, the \(F_s\) used refers to \(F_s'\). \\
Then an lmFH symbol with a frequency shift \(F_s\) and a phase offset \(P_o\) can be expressed as:
\begin{equation} \label{eq:lmFH_Fs_Po}
L_{-u}(k) = \exp(i(2\pi \frac{\sum_{t=0}^{k} (-u \cdot t + F_s)}{P} + P_o)) 
\end{equation}
On the other hand, Eqs. 4 and 12 in Ref. \cite{3} demonstrate that the DFT of a ZC sequences with prime length \(P\) and root \(u\) can be formulated as the conjugate and frequency shift of a dual ZC sequences with root \(u^{-1}\), scaled by a constant factor:
\begin{equation} \label{eq:DFT_Beyme}
F_u(k) = Z_{u^{-1}}^*(k) \cdot e^{i2\pi \frac{2^{-1}(1-u^{-1}) k}{P}} \cdot F_u(0)
\end{equation} 
where \(u^{-1}\) is the inverse of \(u\) modulo \(P\), i.e., \(u\cdot u^{-1}\equiv1 \pmod P\), \(2^{-1}\) is the inverse of \(2\) modulo \(P\), equals to \(\frac{P+1}{2}\), \(^*\) means conjugate, \(Z_{u^{-1}}^*(k) = e^{i\pi \frac{u^{-1}k(k+1)}{P}}\), \(F_u\) is the DFT of \(Z_u\), and \(F_u(0) = \sum_{n=0}^{P-1}Z_u(n)\). \\
According to Eq. \ref{eq:lmFH_Fs}, it can be expressed in the form of lmFH symbols with a frequency shift as:
\begin{equation} \label{eq:DFT_lmFH}
F_{u}(k) = \exp(i 2\pi \frac{\sum_{t=0}^{k} (u^{-1} \cdot t + \frac{P+1}{2}(1-u^{-1}))}{P}) \cdot F_u(0)
\end{equation}
And, according to Eq. \ref{eq:lmFH_Fs_Po} it can be expressed with a phase offset as:
\begin{equation} \label{eq:DFT_lmFH_x}
F_{u}(k) = \exp(i (2\pi \frac{\sum_{t=0}^{k} (u^{-1} \cdot t + \frac{P+1}{2}(1-u^{-1}))}{P} + \angle{F_u(0)})) \cdot \vert F_u(0)  \vert
\end{equation}
Or, further expressed as:
\begin{equation} \label{eq:DFT_lmFH_ZC}
F_{u}(k) = \exp(-i(2\pi \frac{\sum_{t=0}^{k} (-u^{-1} \cdot t + \frac{P+1}{2}(u^{-1}-1))}{P} - \angle{F_u(0)})) \cdot \vert F_u(0) \vert 
\end{equation}
Thus, if we can obtain the argument \(\angle{F_u(0)}\) and magnitude \(\vert F_u(0) \vert\) of \(F_u(0)\), we can rapidly derive the DFT of ZC sequences using Eq. \ref{eq:DFT_lmFH_ZC} based on the principle of mFH. \\
Before delving into this, we first present an intuitive and visual exposition of the DFT of ZC sequences using lmFH patterns.
\subsection{Visual DFT of ZC sequences using lmFH pattern}
\begin{figure}[H] 
	\centering
	\includegraphics[width=7cm]{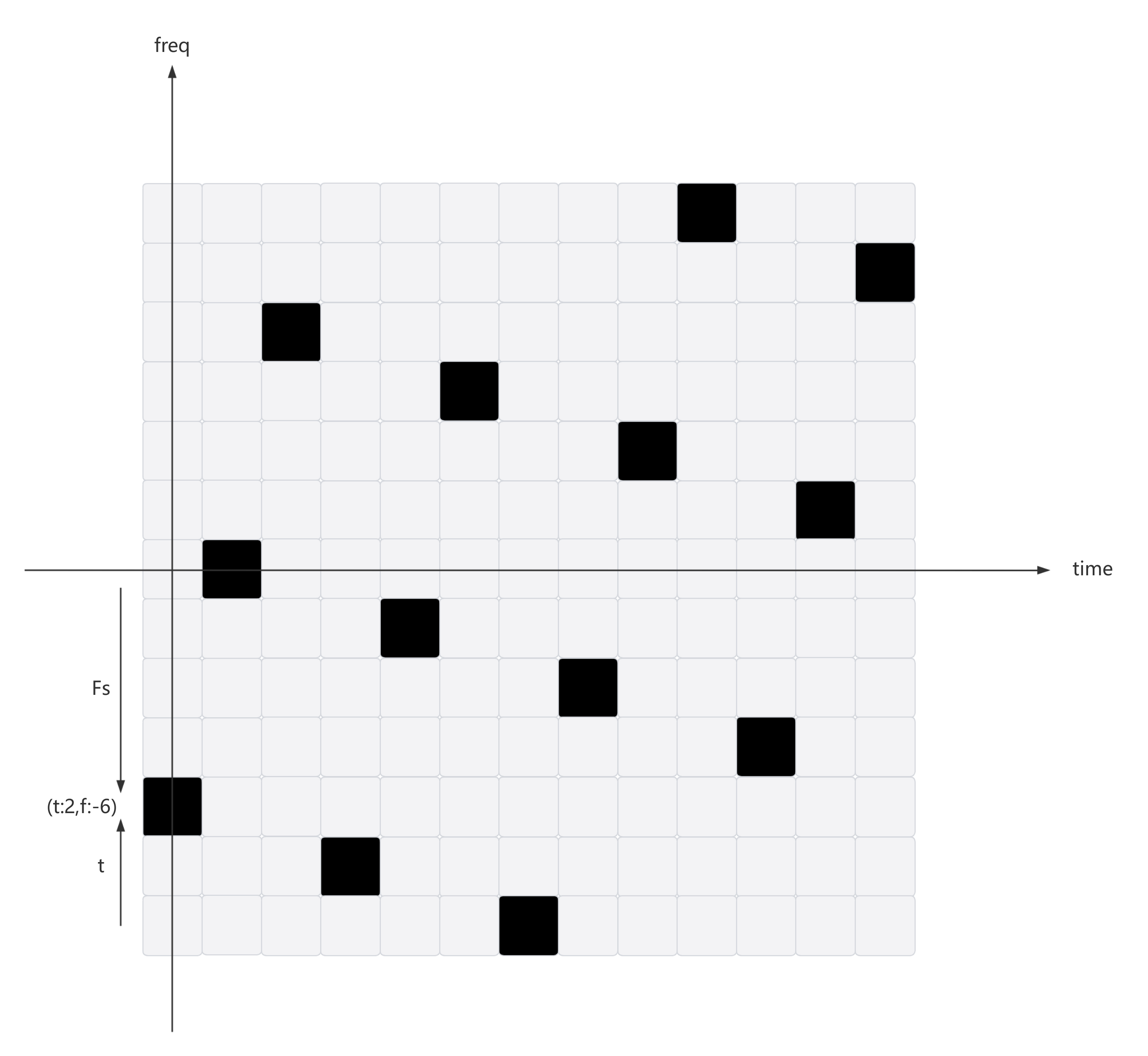}
	\caption{DFT of ZC (reverse side) with P13 \& u3 }
	\label{fg:DFT_P13_u3_reverse}
\end{figure}

\begin{figure}[H] 
	\centering
	\includegraphics[width=7cm]{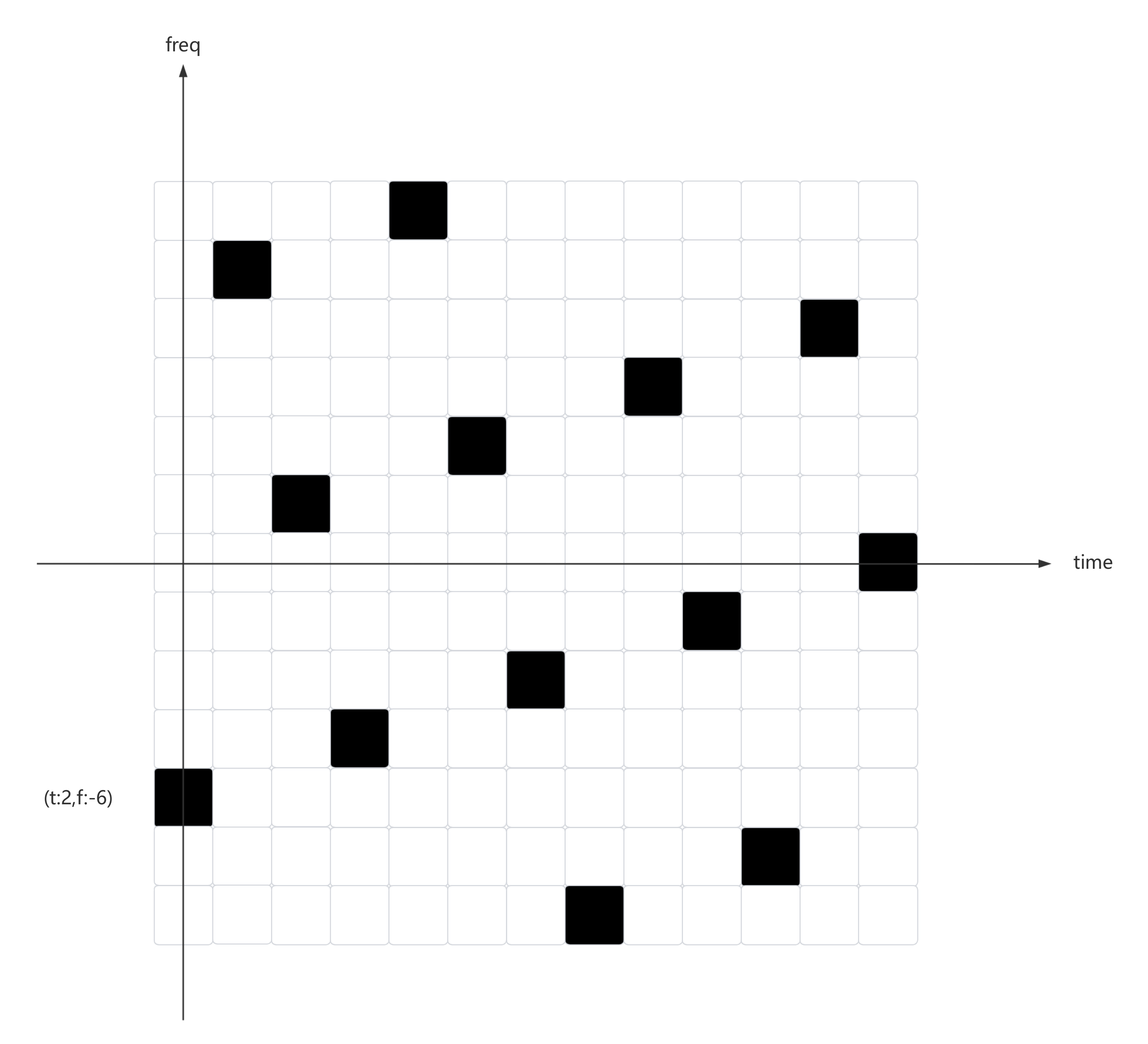}
	\caption{DFT of ZC (obverse side) with P13 \& u3}
	\label{fg:DFT_P13_u3_obverse}
\end{figure}

In this subsection, we use the lmFH pattern to intuitively and visually demonstrate why the DFT of ZC sequences in Eq. \ref{eq:DFT_Beyme} is associated with the conjugate of a dual ZC sequences whose root is the inverse, along with a frequency shift. \\
In the first step, we illustrate the DFT of ZC sequences by flipping the lmFH pattern across the line \(f = t\), thereby transforming the time domain to the frequency domain, as shown in Fig. \ref{fg:DFT_P13_u3_reverse}. In a time-frequency plane, flipping a point \((t, f)\) across \(f = t\) yields the coordinates \((f, t)\), formalized by the matrix equation:
\[
\begin{pmatrix} t \\ f \end{pmatrix}  
\to 
\begin{pmatrix} f \\ t \end{pmatrix} 
= 
\underbrace{\begin{pmatrix} 0 & 1 \\ 1 & 0 \end{pmatrix}}_{\sigma_x}
\begin{pmatrix} t \\ f \end{pmatrix}
\]
The flipping matrix has a determinant of \(-1\), indicating that the "flip" not only transposes the lmFH pattern by swapping \(t\) and \(f\), but also reverses the side of the lmFH pattern due to the determinant's sign. \\
The transposition of the lmFH pattern causes its slope to transform from \(-u\propto\frac{f}{t}\) to its inverse \(-u^{-1}\propto\frac{t}{f}\), while the side reversal induces a right-handed to left-handed coordinate system change, corresponding to complex conjugation. These explain why the DFT of \(Z_u\) is associated with the conjugate of \(Z_{u^{-1}}\), as shown by \(\mathcal{F}\{Z_u\} \propto Z_{u^{-1}}^*\). \\
Interestingly, as \textbf{an extra mathematical result}, Fig. \ref{fg:lmFH_P13_u3} and Fig. \ref{fg:DFT_P13_u3_reverse} present a novel visualization method for obtaining multiplicative inverses, which maps the algebraic inverse operation to flipping the lmFH pattern in the time-frequency plane. \\
Subsequently, we further analyze the origin of frequency shift using Fig. \ref{fg:DFT_P13_u3_reverse} and Fig. \ref{fg:lmFH_P13_u3}. \\
Obviously, the first frequency point in Fig. \ref{fg:DFT_P13_u3_reverse} is not at the zero frequency point. Since the lmFH pattern is obtained by flipping Fig. \ref{fg:lmFH_P13_u3} across the line \(f = t\), the first frequency point in Fig. \ref{fg:DFT_P13_u3_reverse} corresponds to the most negative frequency point in Fig. \ref{fg:lmFH_P13_u3}, whose time value \(t\) satisfies \(t \cdot -u \equiv -\frac{P-1}{2} \pmod{P}\). By solving the congruence relation \(t \cdot u \equiv \frac{P-1}{2} \pmod{P}\), we derive the explicit solution for \(t\), \(t \equiv \frac{P-1}{2}u^{-1} \pmod{P}\). \\
The value \(t\) equals the frequency points difference between the first frequency point and the most negative frequency point in Fig. \ref{fg:DFT_P13_u3_reverse}. Thus, the frequency shift \(F_s\) of the first frequency point is given by: \(F_s = t +(-\frac{P-1}{2})\), further derivation yields: \(F_s \equiv \frac{P-1}{2} (u^{-1} - 1) \pmod{P}\), or \(F_s \equiv \frac{P+1}{2} (1 - u^{-1}) \pmod{P}\) as in Eq. \ref{eq:DFT_Beyme}. \\
Thus, through the first step of flipping the lmFH pattern across the line \(f = t\), we can intuitively and visually explain why the DFT of the ZC sequences is associated with the conjugate of \(Z_{u^{-1}}\) and a frequency shift \(F_s\), as shown by \(\mathcal{F}\{Z_u(k)\} \propto Z_{u^{-1}}^*(k) \cdot e^{i2\pi \frac{\frac{P+1}{2}(1-u^{-1})k}{P}}\). \\
In the second step, we further flip the lmFH pattern in Fig. \ref{fg:DFT_P13_u3_reverse} across the frequency axis to eliminate conjugation, in order to rewriting Eq. \ref{eq:DFT_Beyme} in the form of lmFH symbols, as shown in Fig. \ref{fg:DFT_P13_u3_obverse}. Notice that the first frequency point on the frequency axis after flipping remains on the frequency axis. \\
After the second flip, the lmFH pattern returns to its original side, indicating that it is no longer in the conjugate form. However, its slope transforms from \(-u^{-1}\propto\frac{t}{f}\) to \(u^{-1}\propto\frac{-t}{f}\). Thus, Eq. \ref{eq:DFT_Beyme} can be rewritten as:
\begin{equation} \label{eq:ZC_DFT}
	F_u(k) = Z_{-u^{-1}}(k) \cdot e^{i2\pi \frac{\frac{P+1}{2} (1 - u^{-1}) k}{P}} \cdot F_u(0)
\end{equation}
It can be further expressed in the form of lmFH symbols as Eq. \ref{eq:DFT_lmFH_ZC}. 
\subsection{Visual IDFT of ZC sequences using lmFH pattern}
\begin{figure}[H] 
	\centering
	\includegraphics[width=7cm]{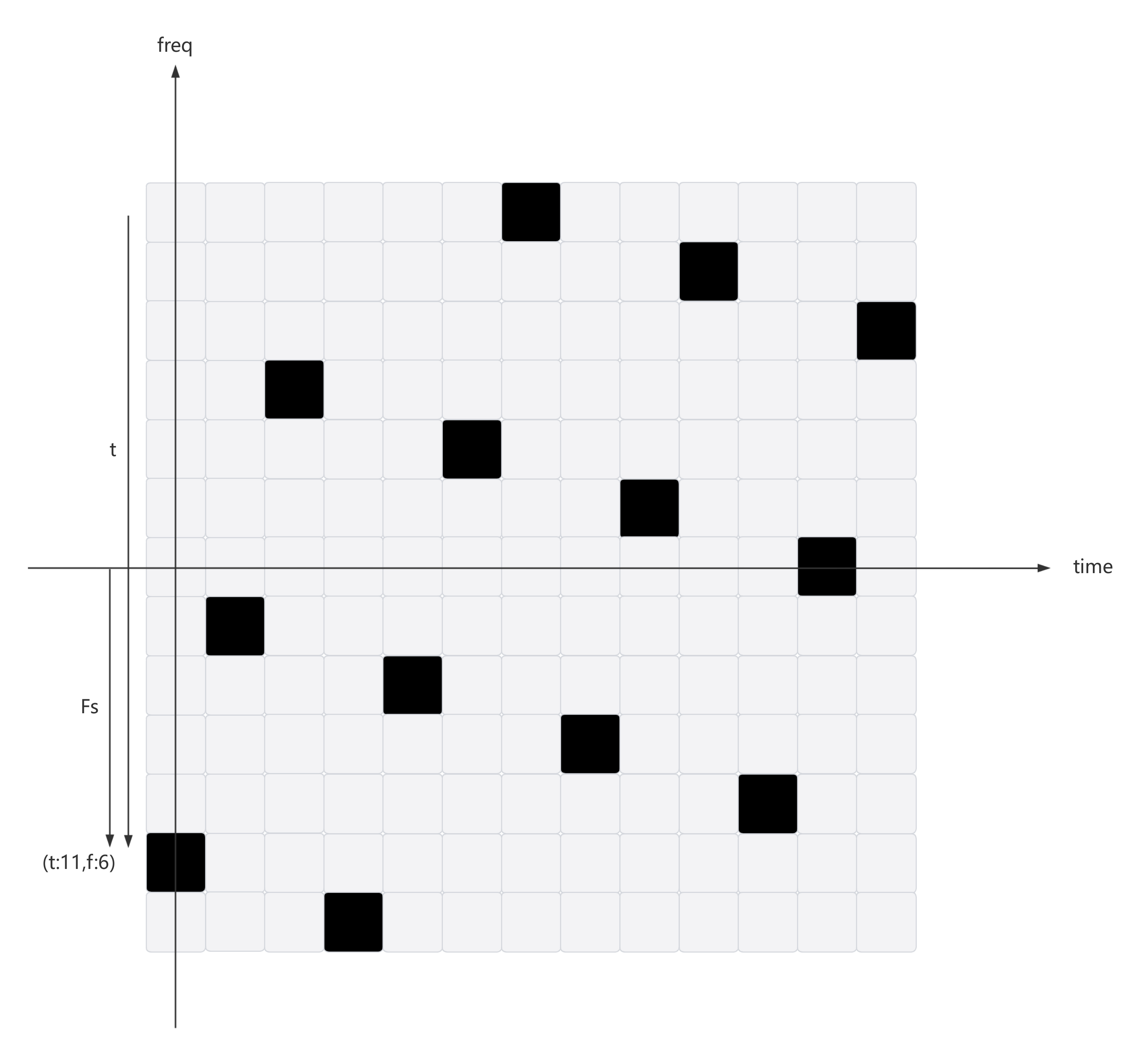}
	\caption{IDFT of ZC (reverse side) with P13 \& u3 }
	\label{fg:IDFT_P13_u3}
\end{figure}

To demonstrate that the operations in the previous section are not coincidental, this section presents an intuitive and visual example on IDFT of ZC sequences. 
The IDFT operation of ZC sequences differs from the DFT operation, it can be achieved by flipping the lmFH pattern across the line \(y = -t\), formalized by the matrix equation:
\[
\begin{pmatrix} t \\ f \end{pmatrix}  
\to 
\begin{pmatrix} -f \\ -t \end{pmatrix} 
= 
\underbrace{\begin{pmatrix} 0 & -1 \\ -1 & 0 \end{pmatrix}}_{-\sigma_x}
\begin{pmatrix} t \\ f \end{pmatrix}
\]
The slope of the transposed lmFH pattern is derived as \(\frac{-t}{-f} = \frac{t}{f} = \left(\frac{f}{t}\right)^{-1} \propto (-u)^{-1}\), indicating that the slope of the lmFH pattern remains \(-u^{-1}\) after IDFT, consistent with its value after DFT, as illustrated in Fig. \ref{fg:IDFT_P13_u3}. \\
Moreover, the side of the lmFH pattern remains reversed, since the determinant of the flipping matrix remains \(-1\). \\
But now, the first frequency point in Fig. \ref{fg:IDFT_P13_u3} corresponds to the most positive frequency point in Fig. \ref{fg:lmFH_P13_u3}, so time value \(t\) needs change to \(t \cdot -u \equiv \frac{P-1}{2} \pmod{P}\), and \(t \equiv \frac{P+1}{2} u^{-1} \pmod{P}\). Then, from Fig. \ref{fg:IDFT_P13_u3} it can be seen that \(F_s=-(t-\frac{P-1}{2}) \equiv \frac{P-1}{2}(u^{-1}+1) \pmod{P}\). \\
Thus, Eq. \ref{eq:DFT_Beyme} for IDFT of ZC can be modified as: 
\begin{equation} \label{eq:ZC_IDFT}
	F_u(k) = Z_{u^{-1}}^*(k) \cdot e^{i2\pi \frac{\mathbf{\frac{P-1}{2}(u^{-1}+1)}k}{P}} \cdot F_u(0)
\end{equation} 
From the lmFH patterns in Fig. \ref{fg:DFT_P13_u3_reverse} and Fig. \ref{fg:IDFT_P13_u3}, another interesting result emerges: the difference between DFT and IDFT of ZC sequences lies in nothing more than a frequency shift of \(1 \pmod{P}\). \\
Mathematically, the difference between the DFT and IDFT frequency shifts is given by \((\frac{P+1}{2}(1-u^{-1}) - \frac{P-1}{2}(u^{-1}+1)) \pmod{P}\), which simplifies to \(1 \pmod{P}\). \\
Note: For a rigorous style, Eq. \ref{eq:ZC_IDFT} can be simply derived as follows: \(\mathcal{IDFT}(Z_u(k)) = \{\mathcal{DFT}(Z_{-u}(k))\}^* = (Z_{-u^{-1}}^*(k)e^{i2\pi\frac{2^{-1}(1-(-u^{-1}))k}{P}}F_{-u}(0))^* = Z_{u^{-1}}^*(k)e^{-i2\pi\frac{\frac{P+1}{2}(1+u^{-1})k}{P}}F_u(0)=Z_{u^{-1}}^*(k)e^{i2\pi\frac{\frac{P-1}{2}(u^{-1}+1)k}{P}}F_u(0)\).
\subsection{Visual DFT of ZC sequences with cyclic shift}
\begin{figure}[H] 
	\centering
	\includegraphics[width=7cm]{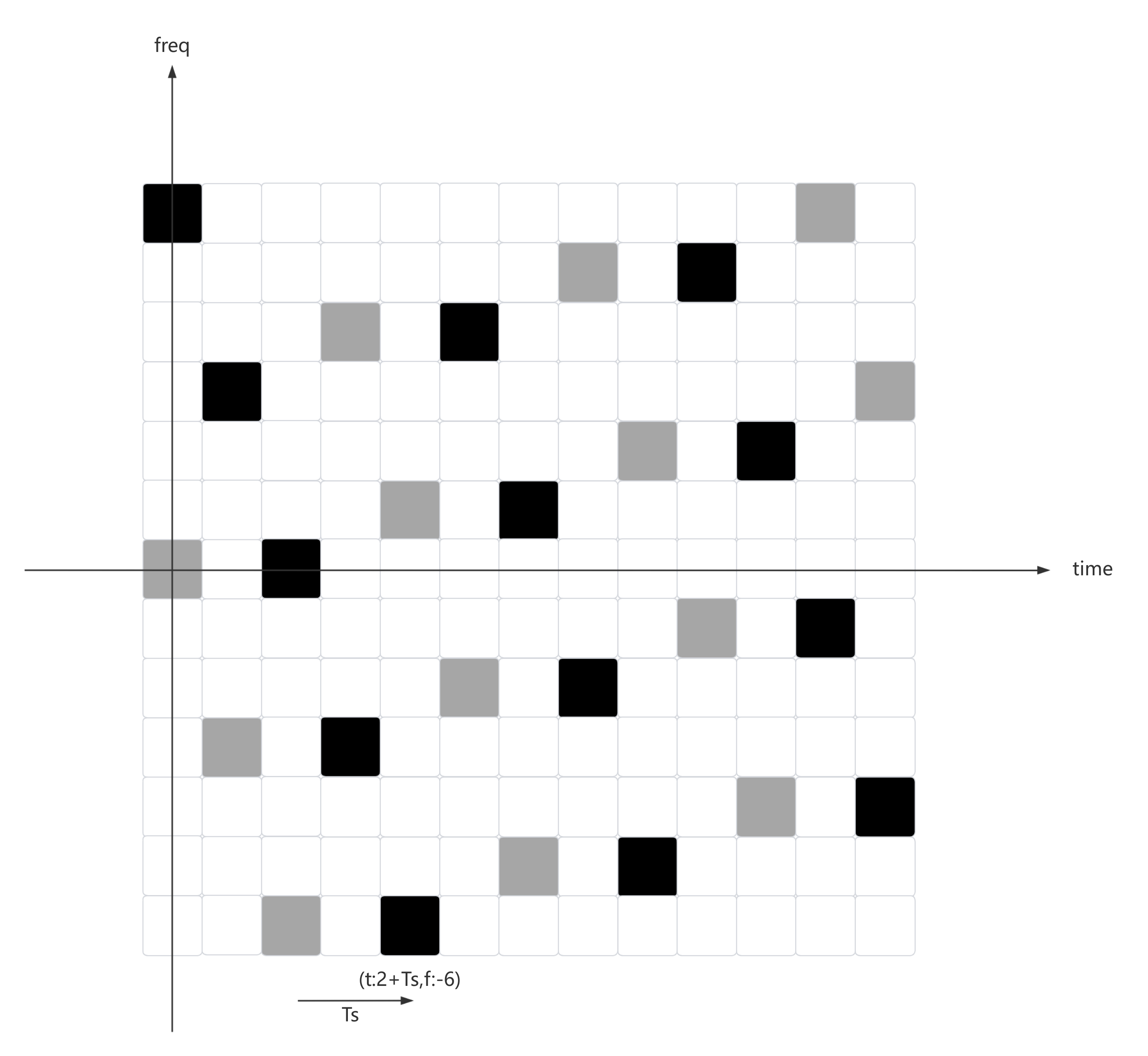}
	\caption{lmFH pattern with cyclic time shift}
	\label{fg:lmFH_TS}
\end{figure}

\begin{figure}[htbp] 
	\centering
	\includegraphics[width=7cm]{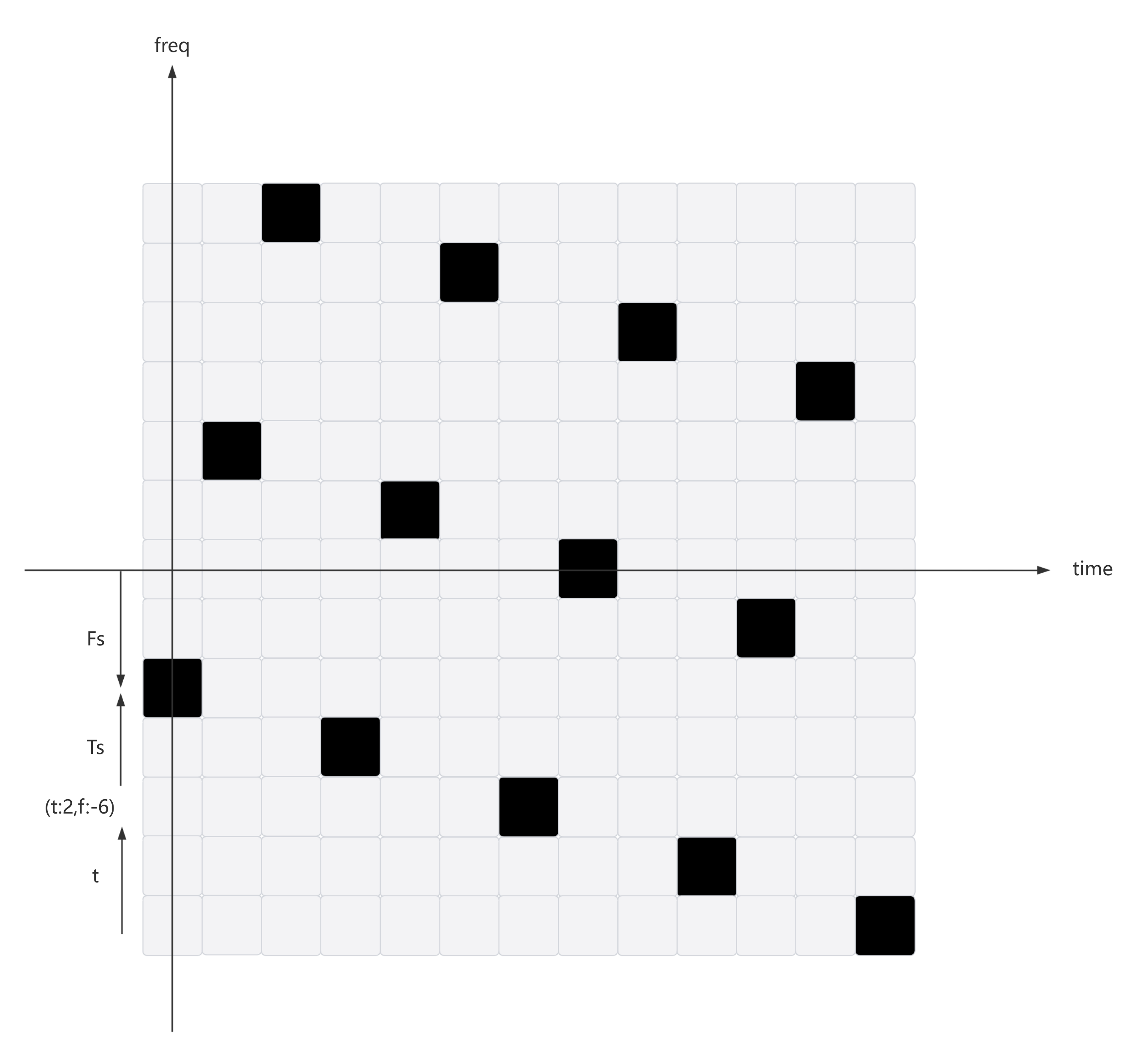}
	\caption{DFT of ZC with cyclic shift (reverse side)}
	\label{fg:DFT_lmFH_TS}
\end{figure}

In scenarios where ZC sequences are actually used in 4G and 5G systems, cyclic shifts play a key role in multi-user differentiation. A ZC sequences with a cyclic shift can be expressed as \(Z_u(k)=e^{-i\pi u\frac{(k+T_s)(k+T_s+1)}{P}}\), where \(T_s\) is the cyclic shift. The "cyclic shift" of ZC sequences is indeed a cyclic time shift of the lmFH pattern, as shown in Fig. \ref{fg:lmFH_TS}. \\
After performing the DFT by flipping the lmFH pattern across \(f = t\), we obtain a frequency domain lmFH pattern of the cyclic shift ZC sequences, as shown in Fig. \ref{fg:DFT_lmFH_TS}. \\
It can be seen in Fig. \ref{fg:DFT_lmFH_TS} that the cyclic time shift of the lmFH pattern in the time domain is equivalent to the cyclic frequency shift in the frequency domain after DFT. So the \(F_s\) in Eq. \ref{eq:ZC_DFT} needs to be changed from \(F_s=\frac{P+1}{2}(1-u^{-1})\) to \(F_s=t+T_s-\frac{P-1}{2}=\frac{P+1}{2}(1-u^{-1})+T_s\). Thus, for the DFT of ZC sequences with cyclic time shift \(T_s\), Eq. \ref{eq:ZC_DFT} can be modified as:
\begin{equation} \label{eq:ZC_DFT_Ts}
		F_u(k) = Z_{-u^{-1}}(k) \cdot e^{i2\pi \frac{(\frac{P+1}{2} (1 - u^{-1})+T_s)k}{P}} \cdot F_u(0)
\end{equation}
Note: A rigorous derivation of Eq. \ref{eq:ZC_DFT_Ts} can be found in Appd. \ref{Ap:ZC_DFT_Ts}. 
\section{Efficient DFT of ZC sequences}
Building on the previous section, we reconstruct Eq. \ref{eq:DFT_Beyme} and extend it to include cyclic shift using the lmFH pattern, as shown in Eq. \ref{eq:ZC_DFT_Ts}. In this section, we first focus on reformulating an efficient method for the cumulative sum of ZC sequences in an alternative form. This enables the transformation of Eq. \ref{eq:ZC_DFT_Ts} into the form of lmFH symbols described by Eq. \ref{eq:DFT_lmFH_ZC}, thereby reducing the complexity of DFT of ZC sequences significantly compared to classical methods.
\subsection{Computing the cumulative sum of ZC sequences using Generalized Quadratic Gauss Sum}
An efficient method for computing the cumulative sum of ZC sequences using the Generalized Quadratic Gauss Sum has been proposed in Proposition 1 of Ref. \cite{4} as shown below:
\begin{equation} \label{eq:GQGS}
	F_u(0) = \sqrt{P} \cdot \ell_{2u} \cdot \eta_P \cdot e^{i2\pi \frac{u(2^{-1})^3}{P}}
\end{equation} 
where \(\ell_{2u}\) denotes the Legendre symbol of \(2u\) with respect to \(P\),
\[
\ell_{2u} = \left(\frac{2^{-1}u}{P}\right) = \left(\frac{2u}{P}\right) = 
\begin{cases} 
	0 & \text{if } P \text{ divides } 2u \\
	1 & \text{if } 2u \not\equiv 0 \text{ is a quadratic residue modulo }P \\
	-1 & \text{if } 2u \not\equiv 0 \text{ is a non-quadratic residue modulo } P
\end{cases}
\]  
And we introduce the coefficient \(\eta_{P}\), 
\[
\eta_{P} = 
\begin{cases} 
	1 & \text{if } P \equiv 1 \pmod{4} \\
	-i & \text{if } P \equiv 3 \pmod{4}
\end{cases}
\]
It can be further expressed in terms of phase:
\begin{equation} \label{eq:GQGS_lmFH}
	\begin{split} 
	F_u(0) 
	&= \sqrt{P} \cdot \exp\left(i2\pi \frac{1-\ell_{2u}}{4}\right) \cdot \exp\left(i2\pi\frac{1-\text{mod}(P,4)}{8}\right) \cdot \exp\left(i2\pi \frac{u(P+1)^3}{8P}\right) \\
	&= \sqrt{P} \cdot \exp\left(i2\pi \left(\frac{1-\ell_{2u}}{4} + \frac{1-\text{mod}(P,4)}{8} + \frac{u(P+1)^3}{8P} \right) \right) \\
	&= \sqrt{P} \cdot \exp\left(i2\pi \frac{(3-2\ell_{2u} - \text{mod}(P,4))P + u(P+1)^3}{8P} \right) \\
\end{split}
\end{equation}
where \(\text{mod}(P,4)\) denotes \(P\) modulo \(4\). Since \(u \not\equiv 0 \pmod{P}\) and \(P\) is prime, \(\ell_{2u} \neq 0\). \\
Let the quasi phase offset be \(QP_o = \frac{(3-2\ell_{2u} - \text{mod}(P,4))P + u(P+1)^3}{8}\), then the magnitude \(\vert F_u(0) \vert\) in Eq. \ref{eq:DFT_lmFH_ZC} is \(\sqrt{P}\), and the argument \(\angle{F_u(0)}\) in Eq. \ref{eq:DFT_lmFH_ZC} is \(2\pi\frac{QP_o}{P}\). Let \(F_s = \frac{P+1}{2}(u^{-1}-1)-T_s\), then Eq. \ref{eq:DFT_lmFH_ZC} can be finally expressed as:
\begin{equation} \label{eq:DFT_ZC_lmFH}
	F_{u}(k) = \exp\left(-i2\pi\frac{\sum_{t=0}^{k} (-u^{-1} \cdot t + F_s)-QP_o}{P}\right) \cdot \sqrt{P} 
\end{equation}
\subsection{Efficient DFT of ZC sequences using lmFH symbols form}
\begin{figure}[H] 
	\centering
	\includegraphics[width=13cm]{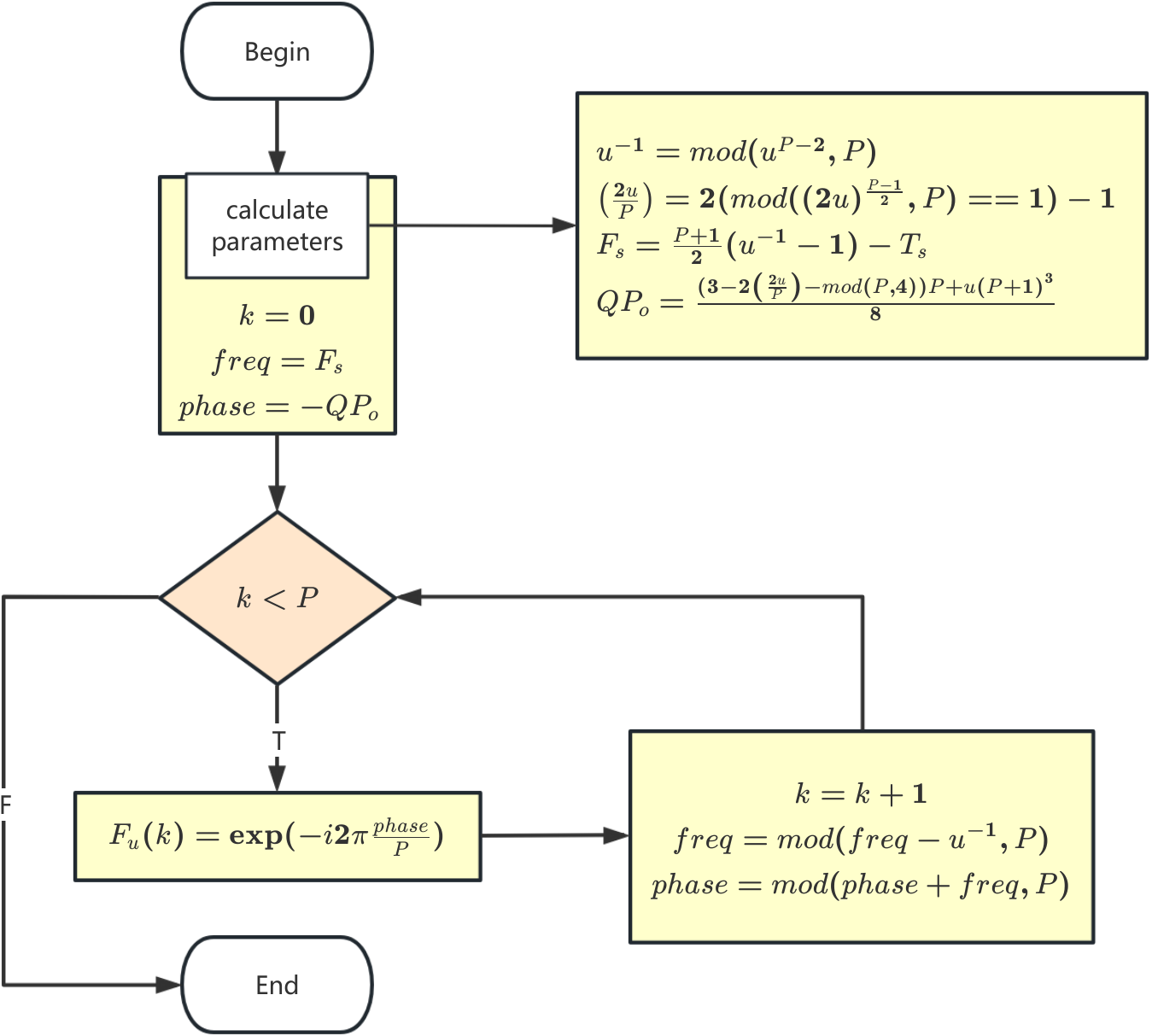}
	\caption{Flowchart of DFT of ZC sequences}
	\label{fg:CHART}
\end{figure}

Inspired by mFH theory and applying the Generalized Quadratic Gauss Sum, we derived an efficient algorithm for the DFT of ZC sequences as shown in Eq. \ref{eq:DFT_ZC_lmFH}, building upon Eqs. \ref{eq:GQGS_lmFH} and \ref{eq:DFT_lmFH_ZC}. \\
Ignoring the magnitude, it can be expressed in pseudocode as Algo. \ref{algo:DFT}: \\
\begin{algorithm} [H] 
	\caption{Efficient DFT of ZC Sequences using lmFH Pattern}
	\label{algo:DFT}
	\textbf{Input:} Prime length \( P \), roots \( u \in \{1, 2, \ldots, P-1\} \), cyclic shift \(T_s\)\\
	\textbf{Output:} DFT coefficients \( F_u(0:P-1) \)
	\begin{algorithmic}[1]
		\State Compute modular inverse: \( iu \gets \text{modInverse}(u, P) \)
		\State Compute Legendre symbol: \( l \gets \text{LegendreSymbol}(2u, P) \)
		\State Compute parameters:
		\begin{align*}
			F_s &\gets \frac{P+1}{2}(iu - 1)-T_s \\
			QP_o &\gets \frac{(3 - 2l - \text{mod}(P,4))P + u(P+1)^3}{8}
		\end{align*}
		\State Initialize phase and frequency:
		\[
		phase \gets -QP_o, \quad freq \gets F_s
		\]
		\For{\( k = 0 \) \textbf{to} \( P-1 \)}
		\State Compute DFT coefficient: \( F_u(k) \gets \exp\left(-i2\pi\frac{ \text{phase}}{P}\right) \)
		\State Update frequency: \( freq \gets freq - iu \mod P\)
		\State Update phase: \( phase \gets phase + freq \mod P\)
		\EndFor
		\State \Return \( F_u \)
	\end{algorithmic}
\end{algorithm}
Or, as shown in the flowchart of Fig. \ref{fg:CHART}:\\ 
Except for some parameters that need to be calculated once, the entire computational load only requires \(2(P-1)\) times additions, \(2(P-1)\) times modulo operations, and \(P\) times complex exponential operations. \\
More specifically, the modular inverses \(u^{-1}\) and Legendre symbols \((\frac{2u}{P})\) can be precomputed and stored in lookup tables. The complex exponential operations can be efficiently implemented using the CORDIC algorithm, where the angles are normalized and multiplied by \(P\), i.e., \(\Theta(i) = \arctan(2^{-i})\frac{P}{2\pi}\). This allows direct computation of the complex exponential using the \(\mathbf{phase}\), eliminating the need for division by \(P\).
\section{Conclusion}
Before the concept of mFH was proposed, the most practical and efficient method for computing the DFT of ZC sequences was Eq. 11 in Ref. \cite{5}. Although numerous patents and papers have proposed improvements, no substantial progress has been made in essence. In this paper, we first reveal a simple and visual relationship between the ZC sequences and the DFT of ZC sequences using lmFH patterns. Then, leveraging the concept of mFH, we transform the computation of the DFT of ZC sequences into the accumulation of frequency points similar to that of normal lmFH symbols. In fact, Algo. \ref{algo:DFT} can be applied to both the DFT and IDFT of ZC sequences with only minor modifications. That is, for DFT of ZC sequences the frequency shift is \(F_s=\frac{P+1}{2}(u^{-1} - 1)-T_s\), and for IDFT of ZC sequences the frequency shift is \(F_s=\frac{P+1}{2}(u^{-1} + 1)+T_s\). 
\section{Acknowledgment}
The author would like to express gratitude to Ms. Pingfang Du for her suggestions on the theoretical derivation of the Generalized Quadratic Gauss Sum.
\bibliographystyle{unsrt}  

\appendix
\begin{center}
	\large\bfseries APPENDIX
\end{center} 
\section{DFT of ZC sequences with cyclic shift} \label{Ap:ZC_DFT_Ts}
According to Eq. 11 in Ref. \cite{5}, the DFT of ZC sequences with cyclic shift \(T_s\) can be expressed as:
\[
	\begin{split} 
		F_u(k) 
		&= Z_u^*(u^{-1}k+T_s)\cdot Z_u(T_s)\cdot F_u(0) \\
		&= e^{i\pi u\frac{(u^{-1}k+T_s)(u^{-1}k+T_s+1)}{P}}\cdot e^{-i\pi u\frac{T_s(T_s+1)}{P}}\cdot F_u(0) \\
		&= e^{i\pi\frac{(k+uT_s)(u^{-1}k+T_s+1)-u(T_s^2+T_s)}{P}}\cdot F_u(0) \\
		&= e^{i\pi\frac{u^{-1}k^2+k(T_s+1)+kT_s}{P}}\cdot F_u(0) \\
		&= e^{i\pi\frac{u^{-1}k^2+2kT_s+k+u^{-1}k-u^{-1}k}{P}}\cdot F_u(0) \\
		&= e^{i\pi u^{-1}\frac{k(k+1)}{P}}\cdot e^{i2\pi\frac{(\frac{P+1}{2}(1-u^{-1})+T_s)k}{P}}\cdot F_u(0) \\
	\end{split}
\] 
\end{document}